\documentclass[twocolumn]{aastex631}
\pdfoutput=1 
\usepackage{amsmath,amstext}
\usepackage[T1]{fontenc}
\usepackage[figure,figure*]{hypcap}
\usepackage[version=4]{mhchem} 

\newcommand{\NH}{{\sl New Horizons}}
\newcommand{\HST}{{\sl HST}}

\shorttitle{Charon's FUV Reflectance}
\shortauthors{Keeney et~al.}

\begin{document}

\title{On Charon's Far-Ultraviolet Surface Reflectance}

\author[0000-0003-0797-5313]{Brian A. Keeney}
\affiliation{Southwest Research Institute, Boulder, CO, USA}
\author[0000-0002-3672-0603]{Joel Wm. Parker}
\affiliation{Southwest Research Institute, Boulder, CO, USA}
\author{Nathaniel Cunningham}
\affiliation{Nebraska Wesleyan University, Lincoln, NE, USA}
\author{S. Alan Stern}
\affiliation{Southwest Research Institute, Boulder, CO, USA}
\author[0000-0002-3323-9304]{Anne J. Verbiscer}
\affiliation{University of Virginia, Charlottesville, VA, USA}
\author{the \NH\ Team}

\correspondingauthor{B. Keeney}
\email{keeney@boulder.swri.edu}

\begin{abstract}
    We present the first measurements of Charon's far-ultraviolet surface reflectance, obtained by the Alice spectrograph on \NH. We find no measurable flux shortward of 1650~\AA, and Charon's geometric albedo is $<0.019$ ($3\sigma$) at 1600~\AA. From 1650--1725~\AA\ Charon's geometric albedo increases to $0.166\pm0.068$, and remains nearly constant until 1850~\AA. As this spectral shape is characteristic of \ce{H2O} ice absorption, Charon is the first Kuiper belt object with a \ce{H2O} ice surface to be detected in the far-ultraviolet. Charon's geometric albedo is $\sim3.7$ times lower than Enceladus' at these wavelengths, but has a very similar spectral shape. We attribute this to similarities in their surface compositions, and the difference in absolute reflectivity to a high concentration or more-absorbing contaminants on Charon's surface. Finally, we find that Charon has different solar phase behavior in the FUV than Enceladus, Mimas, Tethys, and Dione, with a stronger opposition surge than Enceladus and a shallower decline at intermediate solar phase angles than any of these Saturnian satellites.
\end{abstract}

\section{Introduction}
\label{sec:intro}

NASA's \NH\ mission completed its historic flyby of the Pluto system in July 2015 \citep{stern15}. For much of the far-ultraviolet (FUV) Pluto's atmosphere is optically thick \citep{young18}, which complicates efforts to measure its surface reflectance. Nevertheless, \citet{steffl20} found that Pluto has a wavelength-independent surface reflectance ($I/F$) of 0.17 between 1400 and 1850~\AA.

Pluto's FUV reflectance is in the middle of the range found for other icy bodies. For example, the FUV reflectance of Comet~67P/Churyumov-Gerasimenko is only 0.01--0.02 \citep{stern15b}. The FUV reflectance of icy Saturnian satellites varies over a wide range, from $<0.1$ \citep[Phoebe, Iapetus, Dione, Rhea;][]{hendrix08,hendrix08b,hendrix18}, to comparable to Pluto \citep[Mimas, Tethys;][]{hendrix18}, to $>0.3$ \citep[Enceladus;][]{hendrix10,hendrix18}.

Except for Pluto, all of these bodies show the characteristic upturn in reflectance near 1650~\AA\ that is diagnostic of \ce{H2O} ice. Impurities in the \ce{H2O} ice matrix can alter the shape of the upturn \citep[e.g., Ganymede;][]{molyneux20} and different grain sizes can shift the wavelength where the upturn begins \citep{hendrix08}, but neither can produce a constant FUV reflectance between 1400 and 1850~\AA. The brightest regions on Pluto are dominated by \ce{CH4}, \ce{N2}, and \ce{CO} ices \citep{stern15,grundy16}, none of which show an FUV absorption feature. Thus, it is not surprising that Pluto's FUV surface reflectance has a different spectral shape than other icy bodies with surfaces composed mostly of water ice.

Charon's mid-ultraviolet (mid-UV) reflectance was previously measured by the {\sl Hubble Space Telescope} (\HST), finding a nearly constant geometric albedo of $\approx0.25$ from 2250--3300~\AA\  \citep{krasnopolsky01,stern12}. Its FUV reflectance could not be measured due to reduced solar flux at these wavelengths.

 Here we report the first measurements of Charon's FUV surface reflectance, measured by \NH' Alice ultraviolet spectrograph \citep{stern08}. Unlike Pluto, Charon's surface is composed primarily of \ce{H2O} ice \citep{buie87,grundy16}, so its FUV reflectance should more closely resemble that of the icy satellites of the giant planets. We detail the Alice observations of Charon in \autoref{sec:obs}. \autoref{sec:disc} presents Charon's FUV surface reflectance and discusses its implications. We summarize our conclusions in \autoref{sec:conc}.

\begin{deluxetable*}{lcccccccDC}

\tablecaption{Alice Observations of Charon
\label{tab:obs}}

\tabletypesize{\footnotesize}

\tablehead{
  \colhead{Observing} &
  \colhead{UTC Start} &
  \colhead{$t_{\rm exp}$} &
  \colhead{$t_{\rm eff}$} &
  \colhead{$\mathrm{S/N}$\tablenotemark{a}} &
  \colhead{Range} &
  \colhead{Apparent} &
  \colhead{Solar} &
  \twocolhead{Sub-S/C\tablenotemark{b}} & 
  \colhead{$\langle I/F \rangle$\tablenotemark{c}} \\[-1em]
  \colhead{Sequence} &
  \colhead{} &
  \colhead{} &
  \colhead{} &
  \colhead{} &
  \colhead{} &
  \colhead{Diameter} &
  \colhead{Phase} &
  \twocolhead{Lon.} & 
  \colhead{} \\[-1em]
  \colhead{} &
  \colhead{} &
  \colhead{(s)} &
  \colhead{(s)} &
  \colhead{} &
  \colhead{(km)} &
  \colhead{(deg)} &
  \colhead{(deg)} &
  \twocolhead{(deg)} &
  \colhead{}
  }

\decimalcolnumbers
\startdata
PC\_Airglow\_Fill\_2  & 2015-07-13 19:14:11 &  2680 &  2680 & 5.0 &   814,601 & 0.085 & 16.61 &   5.28 & 0.0771\pm0.0016\pm0.0008 \\
PC\_Airglow\_Appr\_1  & 2015-07-13 23:25:11 &   600 &   600 & 3.2 &   625,079 & 0.111 & 17.28 &  -3.06 & 0.0714\pm0.0025\pm0.0022 \\
PC\_Airglow\_Appr\_4  & 2015-07-14 05:20:31 &   600 &   534 & 4.1 &   331,578 & 0.209 & 19.74 & -14.64 & 0.0706\pm0.0018\pm0.0022 \\
C\_LEISA\_LORRI\_1    & 2015-07-14 09:15:23 &   314 &   108 & 2.7 &   140,023 & 0.495 & 26.83 & -17.65 & 0.0623\pm0.0024\pm0.0022
\enddata

\tablenotetext{a}{Median signal-to-noise ratio per pixel (1.8~\AA) in the wavelength range 1650--1850~\AA.}
\tablenotetext{b}{The sub-spacecraft longitude of Charon at the time of observation.}
\tablenotetext{c}{Variance-weighted mean $I/F$ of Charon from 1750--1850~\AA; the first uncertainty is the random uncertainty and the second is the systematic uncertainty (see \autoref{sec:disc:phase}).}

\end{deluxetable*}

\begin{figure*}
  \fig{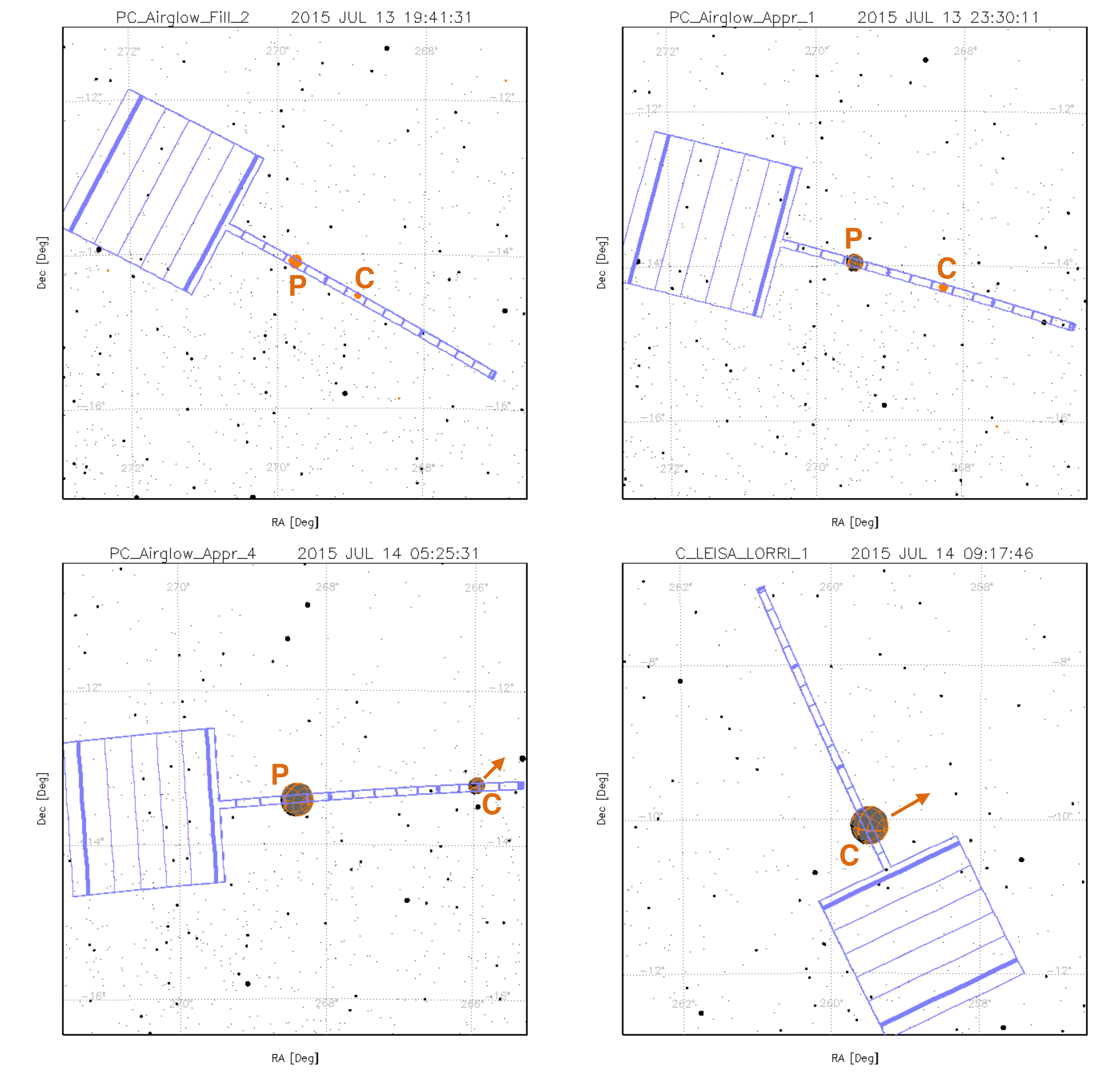}{0.75\textwidth}{}
  \vspace{-2em}
  \caption{The positions of Pluto (``P'') and Charon (``C'') in the Alice slit during the observing sequences in \autoref{tab:obs}. Note that only Charon appears in C\_LEISA\_LORRI\_1, which was designed as a LEISA spectral scan of Charon with ride-along LORRI imagery. The other three sequences were designed to search for airglow emission from Pluto's atmosphere, but the spacecraft was purposely rotated to include Charon in the Alice slit. Arrows show the direction of Charon's motion relative to the slit for sequences where $t_{\rm eff} < t_{\rm exp}$, and all panels have a $6\degr\times6\degr$ field of view.
    \label{fig:gv}}
\end{figure*}

\section{Observations}
\label{sec:obs}

The Alice instrument aboard \NH\ is a lightweight, low-power spectrograph with a bandpass of 520--1870~\AA\ \citep{stern08}. Its imaging microchannel plate detector produces a $1024\times32$ data array, but only the central $780\times21$ pixels are illuminated. This study uses the narrow ($0\fdg1\times4\degr$) portion of the Alice entrance slit, which covers the bottom two-thirds of the detector. Each detector row subtends $0\fdg3$ in the spatial dimension, and the filled-slit spectral resolution is 9~\AA.

\autoref{tab:obs} summarizes the Alice observations of Charon, listing: (1) observing sequence identifier; (2) UTC start time of the exposures analyzed; (3) total exposure time of the exposures analyzed; (4) effective exposure time (i.e., the total time Charon was in the Alice slit) of the exposures analyzed; (5) median signal-to-noise ratio per pixel for Charon in the wavelength range 1650--1850~\AA; (6) \NH' range to Charon;  (7) Charon's apparent angular diameter; (8) Charon's solar phase angle; (9) Charon's sub-spacecraft longitude; and (10) Charon's variance-weighted mean $I/F$ from 1750--1850~\AA\ and its random and systematic uncertainties (see \autoref{sec:disc:phase}). These observations were obtained as part of longer ``observing sequences,'' or collections of exposures obtained for the same scientific purpose. Within each sequence, we focus our analysis on the subset of exposures where Charon's position in the Alice slit is relatively stable (i.e., there is minimal jitter), and only details of those exposures are tabulated.
 
\autoref{fig:gv} shows the positions of Pluto and Charon in the Alice slit for each observing sequence. These snapshots are calculated at the midpoint of each observing sequence, or the midpoint of Charon's scan through the slit if $t_{\rm eff} < t_{\rm exp}$, as are the Charon-specific values in \autoref{tab:obs}. We analyze nine exposures in PC\_Airglow\_Fill\_2, and it is the only sequence in which Charon's angular diameter is smaller than the Alice slit width. PC\_Airglow\_Appr\_1 consists of two 300-s exposures and the pointing is stable throughout. Charon's position in the slit is also stable during the first three 150-s exposures of PC\_Airglow\_Appr\_4, but it drifts out of the slit partway through the final exposure. Finally, C\_LEISA\_LORRI\_1 is a single integration where Charon scans through the slit mid-exposure. The effective exposure time for PC\_Airglow\_Appr\_4 and C\_LEISA\_LORRI\_1 is the amount of time that Charon was in the slit during the exposures analyzed, as determined by the GeoViz\footnote{\url{https://geoviz.space.swri.edu/}} software package \citep{throop09} using SPICE kernels \citep{acton18} created from reconstructed instrument pointing.

\begin{figure}
  \fig{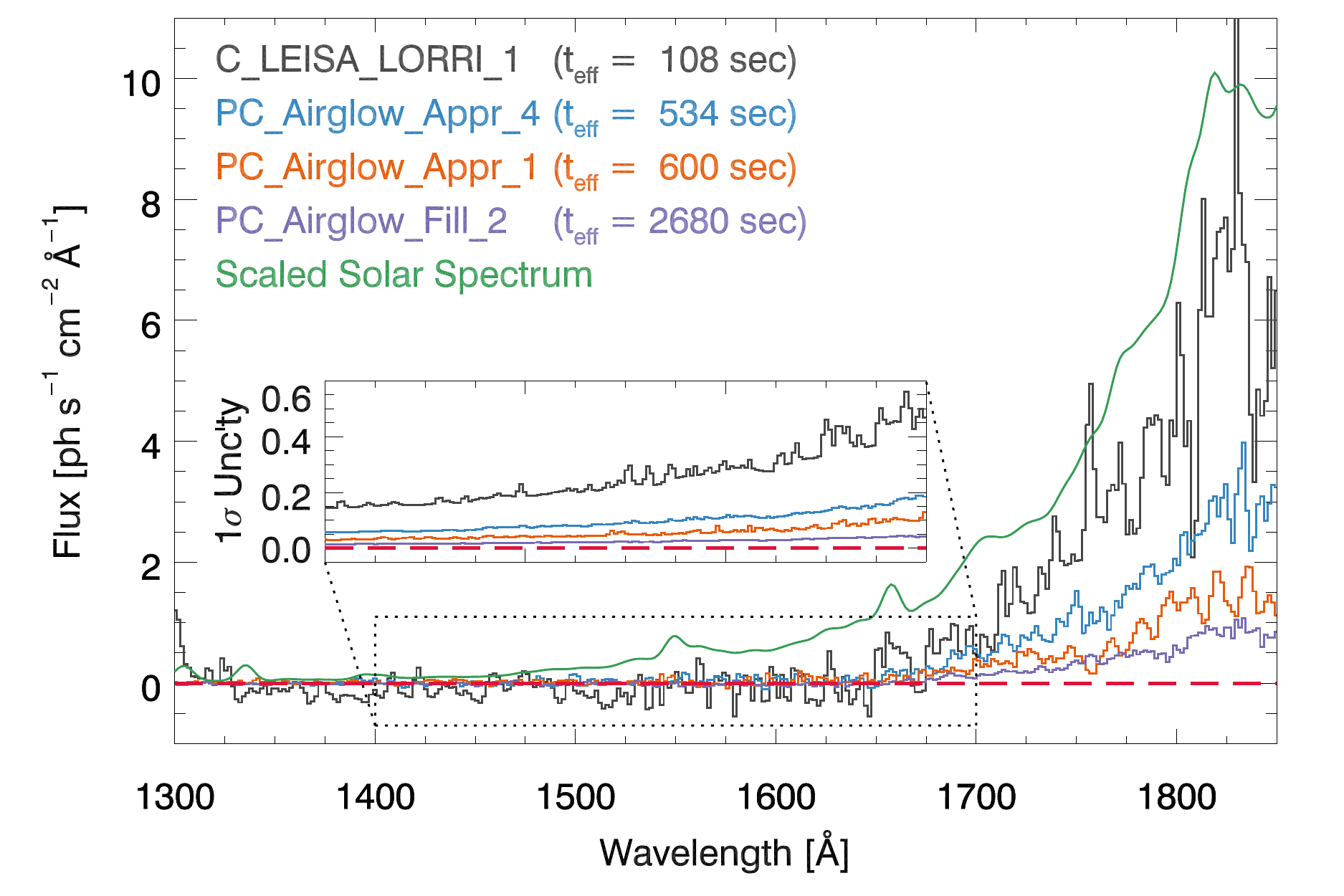}{\columnwidth}{}
  \vspace{-3em}
  \caption{FUV spectra of Charon extracted from the observing sequences in \autoref{tab:obs}. A scaled solar spectrum is shown for comparison. The inset shows the $1\sigma$ flux uncertainty for each observing sequence from 1400--1700~\AA.
    \label{fig:spec}}
\end{figure}

All exposures were corrected as described in \citet{steffl20} for the effects of detector dead time, dark counts, and scattered light from the wings of the Ly$\alpha$ profile generated by hydrogen in the interplanetary medium. Spectra of Charon were then extracted from individual exposures, and when possible exposures from the same observing sequence were combined using a variance-weighted mean. These coadditions are shown in \autoref{fig:spec}, along with the solar spectrum at Pluto arbitrarily reduced to fit on the scale of the plot. We adopt the solar spectrum of \citet{young18}, which combines SUMER reference spectra \citep{curdt01} with observations from TIMED/SEE \citep{woods05} for the \NH\ observation dates.

Owing largely to declining solar flux, Charon shows no observable flux below 1650~\AA\ in any of the sequences, and the observed flux at longer wavelengths increases as \NH\ approaches Charon. However, Charon's position in the Alice slit is most stable (i.e., its $t_{\rm eff}$ is largest) when \NH\ is far from Charon. The combination of these effects leads to the modest and relatively constant signal-to-noise ratios in \autoref{tab:obs}.

\section{Results and Discussion}
\label{sec:disc}

We determine the radiance of Charon in each observing sequence by dividing the fluxes in \autoref{fig:spec} by Charon's solid angle as observed by Alice. We then take the ratio of Charon's radiance to the radiance of a Lambertian surface normally illuminated by the Sun (i.e., $F_{\Sun}/\pi$) to derive Charon's surface reflectance ($I/F$). The spectral shape of Charon's $I/F$ is the same in all sequences, and the inset of \autoref{fig:phase} shows the surface reflectance from PC\_Airglow\_Appr\_4 as an example. The characteristic upturn from \ce{H2O} ice absorption is present, as the reflectance rapidly increases from zero blueward of 1650~\AA\ to 0.07 at 1725~\AA, and remains at that level until 1850~\AA.

\subsection{Solar Phase Dependence}
\label{sec:disc:phase}

\begin{figure}
  \fig{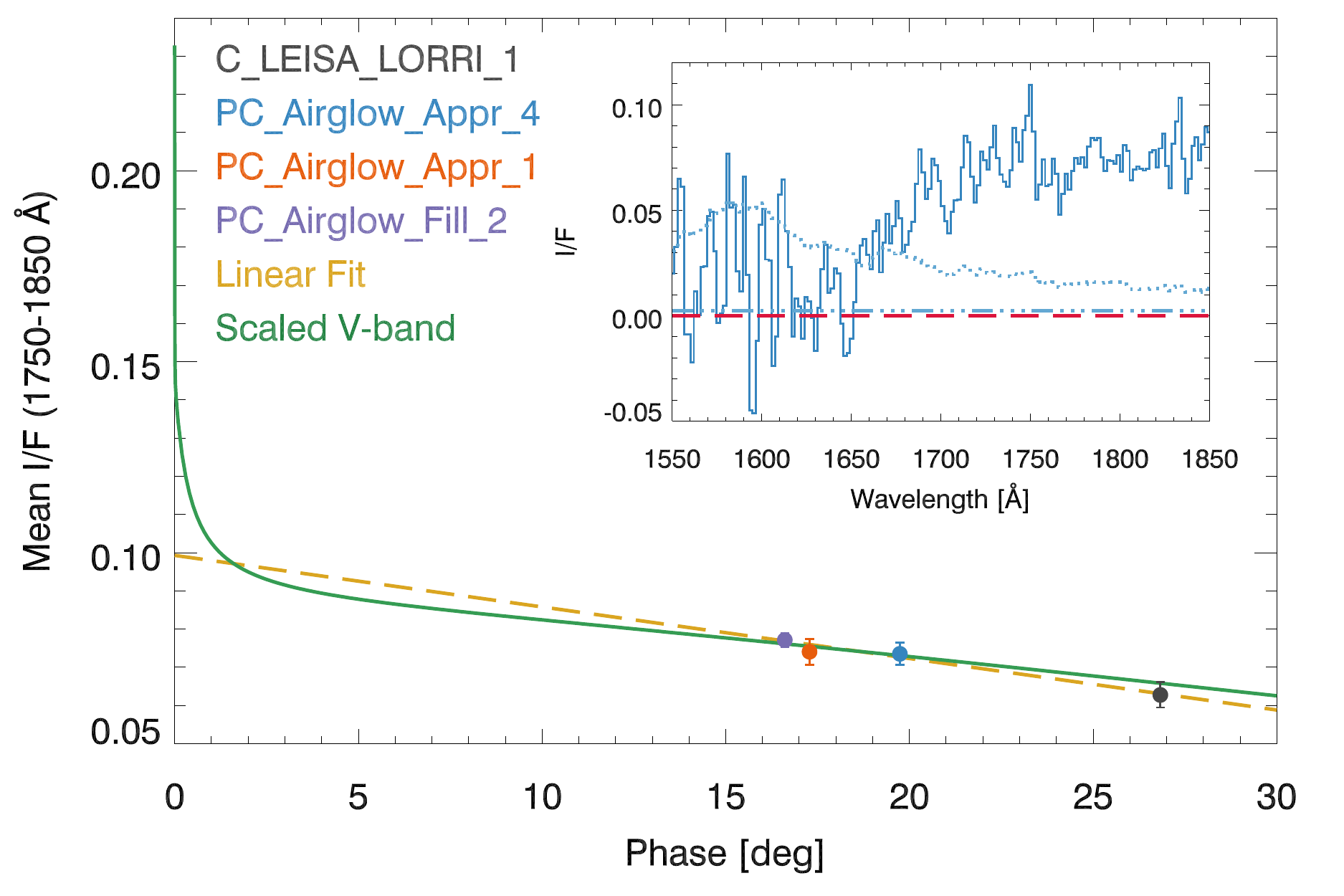}{\columnwidth}{}
  \vspace{-3em}
  \caption{The variance-weighted mean $I/F$ of Charon from 1750--1850~\AA\ as a function of solar phase angle. The dashed line shows a linear fit to the Alice data. The solid line shows Charon's $V$-band phase curve (Verbiscer et~al. 2021, in preparation) scaled to match the Alice data. The inset shows the surface reflectance as a function of wavelength for PC\_Airglow\_Appr\_4. The dotted line shows the $1\sigma$ random uncertainty in $I/F$ and the dot-dashed line shows the $1\sigma$ systematic uncertainty.
    \label{fig:phase}}
\end{figure}

\autoref{fig:phase} shows Charon's variance-weighted mean $I/F$ from 1750--1850~\AA\ as a function of solar phase angle. We estimate two different statistical uncertainties in Charon's surface reflectance. The first is a random uncertainty based on photon-counting (i.e., Poisson) statistics and the second is a systematic uncertainty that accounts for Charon's motion in the Alice slit. We assume a 3\% systematic pointing uncertainty for sequences where Charon fills the slit, and a 1\% pointing uncertainty for sequences where it does not. Further, for sequences where $t_{\rm eff} < t_{\rm exp}$ we assume a 2-s uncertainty on $t_{\rm eff}$.

Since Charon's $I/F$ is nearly constant at these wavelengths (see \autoref{fig:phase} inset) and we average over $>50$ spectral pixels, the random uncertainties in the mean $I/F$ are small despite the modest signal-to-noise ratio of the data; however, the systematic uncertainties are not reduced when averaging. The error bars in \autoref{fig:phase} are the quadrature sum of the random and systematic uncertainties for each sequence. Their individual contributions are listed in the final column of \autoref{tab:obs}.

We use the photometric model of (Verbiscer et~al. 2021, in preparation) to study changes in Charon's relative FUV reflectance with solar phase. This model expands upon the models of \citet{buratti19} and \citet{howett21} by fitting \NH\ LORRI images and \HST\ images from Programs 13667 (Buie PI), 15261 (Verbiscer PI), and 15505 (Verbiscer PI). The phase angle ranges of the \citeauthor{buratti19} and \citeauthor{howett21} data sets were $0\fdg0612$--$169\fdg5$.  The phase angle range of the Verbiscer et~al. data set was $0\fdg0049$--$169\fdg5$, and included 10~points between $0\fdg0049$ and $0\fdg0612$ from \HST\ Programs 15261 and 15505 that were not included in the previous analyses. 

The best-fit \citet{hapke12} parameters for the Verbiscer et~al. (2021, in preparation) model are a single-scattering albedo at 5500~\AA\ of $0.70\pm0.12$, two-term \citet{henyey41} phase function parameters of $b=0.250\pm0.095$ and $c=0.45\pm0.38$, macroscopic roughness of $28\fdg0\pm0\fdg5$, shadow-hiding amplitude of $1.0^{+0.0}_{-0.058}$ and width of $0.0035\pm0.161$, and coherent backscatter amplitude of $0.6\pm0.39$ and width of $0.000033\pm0.00257$. A least-squares fit finds that the solar phase curve of Verbiscer et~al. (2021, in preparation) matches the Alice data when a multiplicative scale factor of $0.592\pm0.010$ is applied, as was done to produce the solid line in \autoref{fig:phase}; this is equivalent to the Verbiscer et~al. solar phase curve with a single-scattering albedo at 1800~\AA\ of $0.414\pm0.071$. Once this multiplicative scale factor is applied, the scaled $V$-band model is consistent with the Alice measurements obtained at phase angles of $17\degr$--$27\degr$ (see \autoref{tab:obs}), as is a simple linear fit (\autoref{fig:phase}), but they have very different predictions for Charon's surface reflectance at zero phase. We discuss this in detail in \autoref{sec:disc:albedo}.

Charon's FUV phase coefficient (the slope of the solar phase curve; see dashed line in \autoref{fig:phase}) over the range of solar phase angles probed by Alice is $0.020\pm0.004$~mag/deg. This value is much lower than that of Dione's leading hemisphere \citep[0.039~mag/deg;][]{royer14} measured over the same range in solar phase angle. With the exception of Enceladus, all other icy Saturnian satellites studied by Cassini/UVIS have steeper solar phase curves than Dione's leading hemisphere (see \autoref{sec:disc:enc}). This difference in solar phase behavior may be due to E-ring particle and/or high-energy electron bombardment in the Saturn system that Charon does not experience. It may also result from physical processes unique to Charon; e.g., conversion of seasonally cold-trapped non-water volatiles escaping from Pluto's atmosphere into reddish tholins in Mordor Macula by photolysis/radiolysis \citep{grundy16}.

\subsection{Charon's Partial FUV Rotation Curve}
\label{sec:disc:rot}

\begin{figure}
  \fig{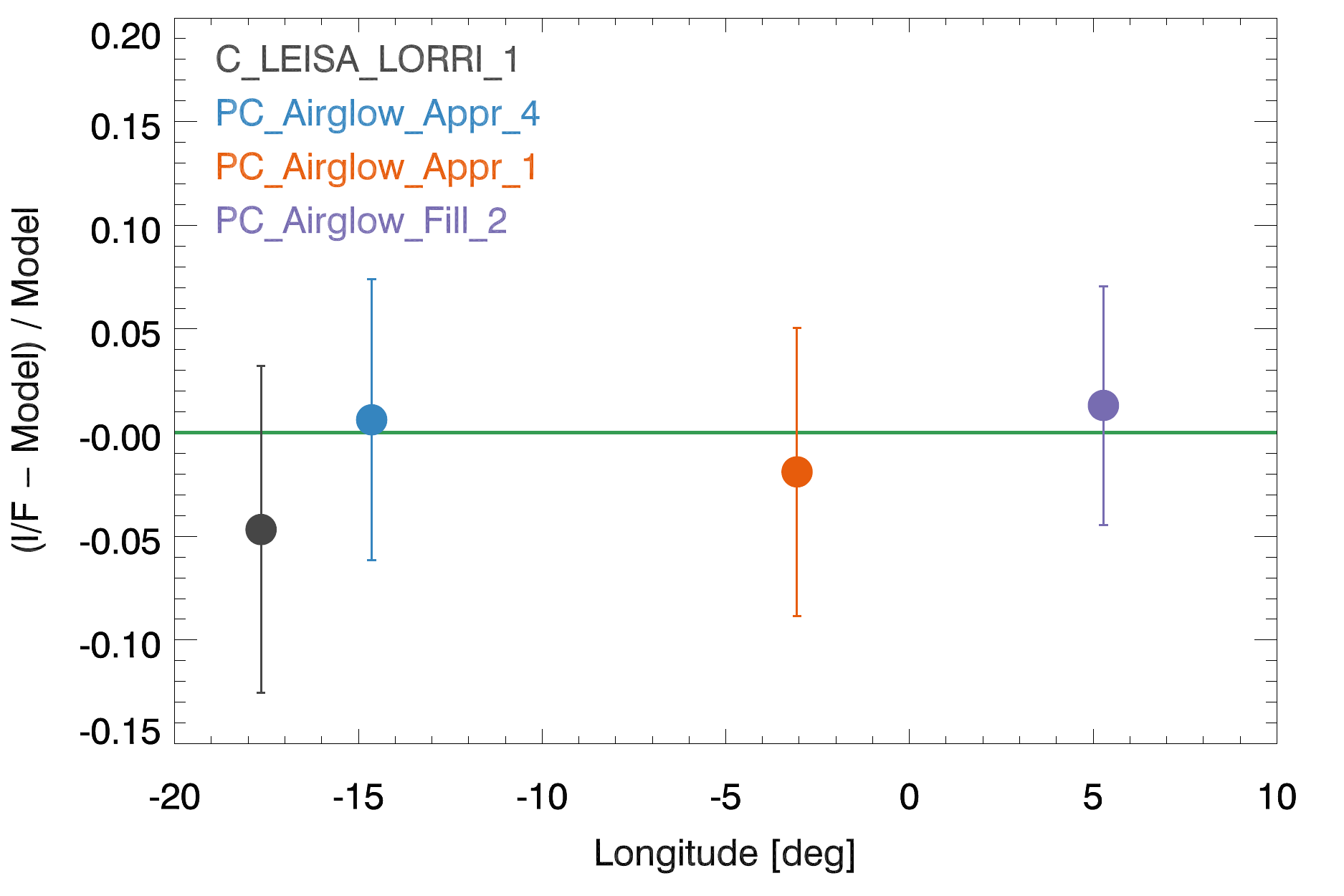}{\columnwidth}{}
  \vspace{-3em}
  \caption{Relative change in Charon's surface reflectance with longitude, where measured. The scaled $V$-band phase curve of \autoref{fig:phase} has been used to model the effects of differing solar phase angles for each observing sequence, and the $1\sigma$ uncertainties displayed take into account the uncertainty in the scale factor.
    \label{fig:longitude}}
\end{figure}

We use the scaled $V$-band phase curve from \autoref{fig:phase} to compensate for the varying solar phase angles of each observing sequence. After this correction is applied we can search for brightness changes in Charon's FUV rotation curve, despite the limited rotational range ($\sim24\degr$ or $\sim1/15$ of a rotation) covered by the Alice observations. The result is shown in \autoref{fig:longitude}.

We find no evidence of variability as a function of longitude in the Alice observing sequences. All of the sequences are consistent with the model expectation, and each other, to within the statistical uncertainties. The maximum absolute deviation from the model expectation is $4.7\pm7.9$\% and the maximum deviation among the Alice sequences is $6.0\pm9.8$\%. These uncertainties take into account the uncertainty in the multiplicative scale factor for the $V$-band phase curve. This result is consistent with Charon's $V$-band rotation curve, which has an amplitude of just 8\% \citep[0.08~mag;][]{buie97,buie10}.

We have also searched for hemispherical differences in Charon's brightness in PC\_Airglow\_Appr\_4 and C\_LEISA\_LORRI\_1, where Charon subtends more than one spatial row on the Alice detector. We find no row-to-row variations in either spectrum at the $>1\sigma$ level. However, the modest signal-to-noise ratio of the data precludes us from detecting variations $\lesssim25$\%.

Although the above limits on Charon's rotational (longitudinal) and hemispherical (latitudinal) FUV surface reflectance variations are not particularly constraining, it will likely be decades before another FUV spectrograph obtains superior data of Charon. Thus, we believe it is important to report the results derived from all the data in hand, even if they are inconclusive.

\newpage
\subsection{Charon's FUV Geometric Albedo}
\label{sec:disc:albedo}

The Alice observing sequences must be corrected to zero phase to determine Charon's FUV geometric albedo. Both the scaled $V$-band solar phase curve of Verbiscer et~al. (2021, in preparation) and a simple linear fit are a good match to the Alice data (\autoref{fig:phase}) at solar phase angles of $17\degr$--$27\degr$. Extrapolating these curves to zero phase yields geometric albedos of $0.2328\pm0.0040$ and $0.0993\pm0.0054$, respectively. 
The random statistical uncertainties of these predictions are much smaller than the variation between them, so the systematic uncertainty of extrapolating to zero phase in the absence of constraining FUV data clearly dominates.

Since it predicts no opposition surge at all, the linear fit to the Alice data sets a lower bound on Charon's geometric albedo at 1800~\AA. The scaled $V$-band model predicts a very strong opposition surge; however, the strength of the opposition surge varies with wavelength and is weaker in the FUV than in $V$-band \citep{hapke21}. Therefore, we treat the scaled $V$-band model as an upper bound to Charon's geometric albedo at 1800~\AA. Even though the strength of the opposition surge is weaker in the FUV than in $V$-band, we nonetheless expect one to be present in the FUV since bodies such as Charon with intermediate albedos are found to have the strongest opposition surges \citep{nelson04,hapke21}. In the discussion below we adopt a value halfway between the bounds discussed above; i.e., a geometric albedo of $0.166\pm0.068$.

\begin{figure}
  \fig{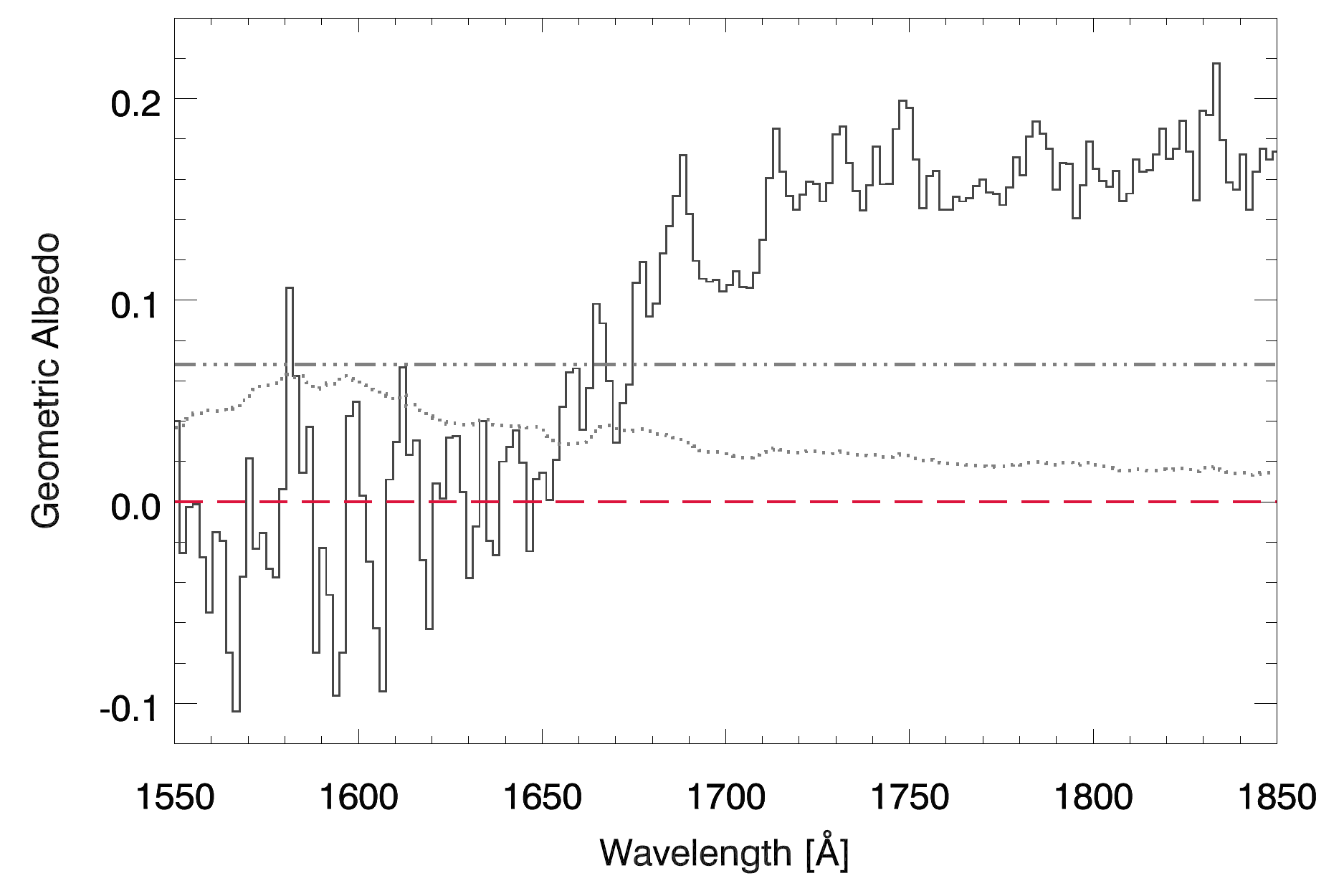}{\columnwidth}{}
  \vspace{-3em}
  \caption{Charon's average geometric albedo as a function of wavelength. The dotted line shows the $1\sigma$ random uncertainty and the dot-dashed line shows the $1\sigma$ systematic uncertainty. 
    \label{fig:albedo}}
\end{figure}

Since there is no evidence of brightness variations in Charon's FUV rotation curve (\autoref{fig:longitude}), after correcting each sequence to zero phase we combine them using a variance-weighted mean to increase the signal-to-noise ratio. \autoref{fig:albedo} shows the geometric albedo of Charon obtained from averaging all of the Alice observing sequences. A least-squares fit to the plateau in the 1710--1850~\AA\ region finds a geometric albedo of $0.1658\pm0.0021$ at 1800~\AA\ and a slope of $63\pm32$\% per 1000~\AA; this slope is only valid for the relatively flat region from 1710-1850~\AA, not the transition region from approximately 1650-1700~\AA. Charon's geometric albedo is $<0.019$ ($3\sigma$) at 1600~\AA, derived from a variance-weighted mean of the measurements from 1550-1650~\AA. The uncertainties on these values only take into account the random uncertainty in Charon's geometric albedo. Its systematic uncertainty is much larger, as discussed above.

Charon's geometric albedo at 1800~\AA\ is lower than the value of $\approx0.25$ from 2250--3000~\AA\ found by \HST\ \citep{krasnopolsky01,stern12}. Although the FUV spectral slope is substantial, it is not large enough to reconcile the measurements, predicting a geometric albedo of $0.213\pm0.026$ at 2250~\AA\ using the linear fit to the 1710--1850~\AA\ region described above. However, this difference is not surprising since Charon is composed primarily of \ce{H2O} ice, like the Saturnian satellites, and \ce{H2O}-ice-rich surfaces decrease strongly in reflectance between 2250~\AA\ and 1850~\AA\ \citep[e.g.,][]{hendrix18}. The scaled $V$-band phase curve of Verbiscer et~al. (2021, in preparation) predicts a geometric albedo at 1800~\AA\ that is much closer to the value measured by \HST\ at 2250~\AA, which we interpret as further evidence that it can be treated as an upper bound to Charon's FUV geometric albedo.

\begin{deluxetable}{llCCC}

\tablecaption{Geometric Albedos of Charon and Some Saturnian Satellites
\label{tab:alb}}

\tablehead{
  \colhead{Satellite} &
  \colhead{Hemisphere} &
  \colhead{1800~\AA} & 
  \colhead{3000~\AA} & 
  \colhead{5500~\AA}
  }

\startdata
Charon    & \nodata  & 0.166\pm0.068  & 0.25\pm0.01^{(1)} & 0.41\pm0.01^{(2)}   \\
Mimas     & Leading  & 0.37^{(3)}     & 0.583^{(4)}       &  0.920\pm0.004^{(5)} \\
          & Trailing & 0.38^{(3)}     & 0.639^{(4)}       & 1.008\pm0.004^{(5)} \\
Enceladus & Leading  & 0.59^{(4)}     & 0.881^{(4)}       & 1.329\pm0.008^{(5)} \\
          & Trailing & 0.63^{(4)}     & 0.944^{(4)}       & 1.425\pm0.008^{(5)} \\
Tethys    & Leading  & 0.46^{(3)}     & 0.868^{(4)}       & 1.288\pm0.005^{(5)} \\
          & Trailing & 0.31^{(3)}     & 0.652^{(4)}       & 1.175\pm0.005^{(5)} \\
Dione     & Leading  & 0.27^{(3)}     & 0.799^{(4)}       & 1.261\pm0.004^{(5)} \\
          & Trailing & 0.14^{(3)}     & 0.424^{(4)}       & 0.826\pm0.004^{(5)} \\
Rhea      & Leading  & \sim0.16^{(4)} & 0.572^{(4)}       & 1.077\pm0.003^{(5)} \\
          & Trailing & \sim0.12^{(4)} & 0.439^{(4)}       & 0.848\pm0.002^{(5)} 
\enddata

\tablerefs{(1) \citealt{stern12}; (2) \citealt{buratti17}; (3) \citealt{royer14}; (4) \citealt{hendrix18}; (5) \citealt{verbiscer07}. \explain{Updated Charon's 1800~\AA\ albedo to match text.}}

\end{deluxetable}

\subsection{Comparisons to Icy Saturnian Satellites}
\label{sec:disc:enc}

Charon's geometric albedos in the FUV, mid-UV, and visible are compared to the albedos of icy Saturnian satellites in \autoref{tab:alb}. Since the Saturnian satellites often show leading-trailing hemispheric brightness differences, we tabulate them separately. All of the satellites in \autoref{tab:alb} are considerably brighter in the visible than the UV; however, the steepness of the decline varies. Between 5500~\AA\ and 3000~\AA, the albedo of all satellites declines by a factor of 1.5--2. The range is  wider between 3000~\AA\ and 1800~\AA, where the albedos of Charon, Mimas, Enceladus, and Tethys decline by a factor of $\lesssim2$, and the albedos of Dione and Rhea decline by a factor of 3--4. Overall, the albedos of these satellites decrease by factors ranging from 2.3 (Enceladus) to $\sim7$ (Rhea) between 5500~\AA\ and 1800~\AA.

Comparing the albedos at each wavelength finds that Mimas (2.2--2.6) and Enceladus (3.2--3.9) have the most consistent albedo ratios compared to Charon. This suggests that their spectral shapes from the FUV to the visible are most similar to Charon's. Rhea and the trailing hemisphere of Dione are the most dissimilar to Charon because their albedos are comparable at 1800~\AA\ but more than two times larger at 5500~\AA.

\begin{figure}
  \fig{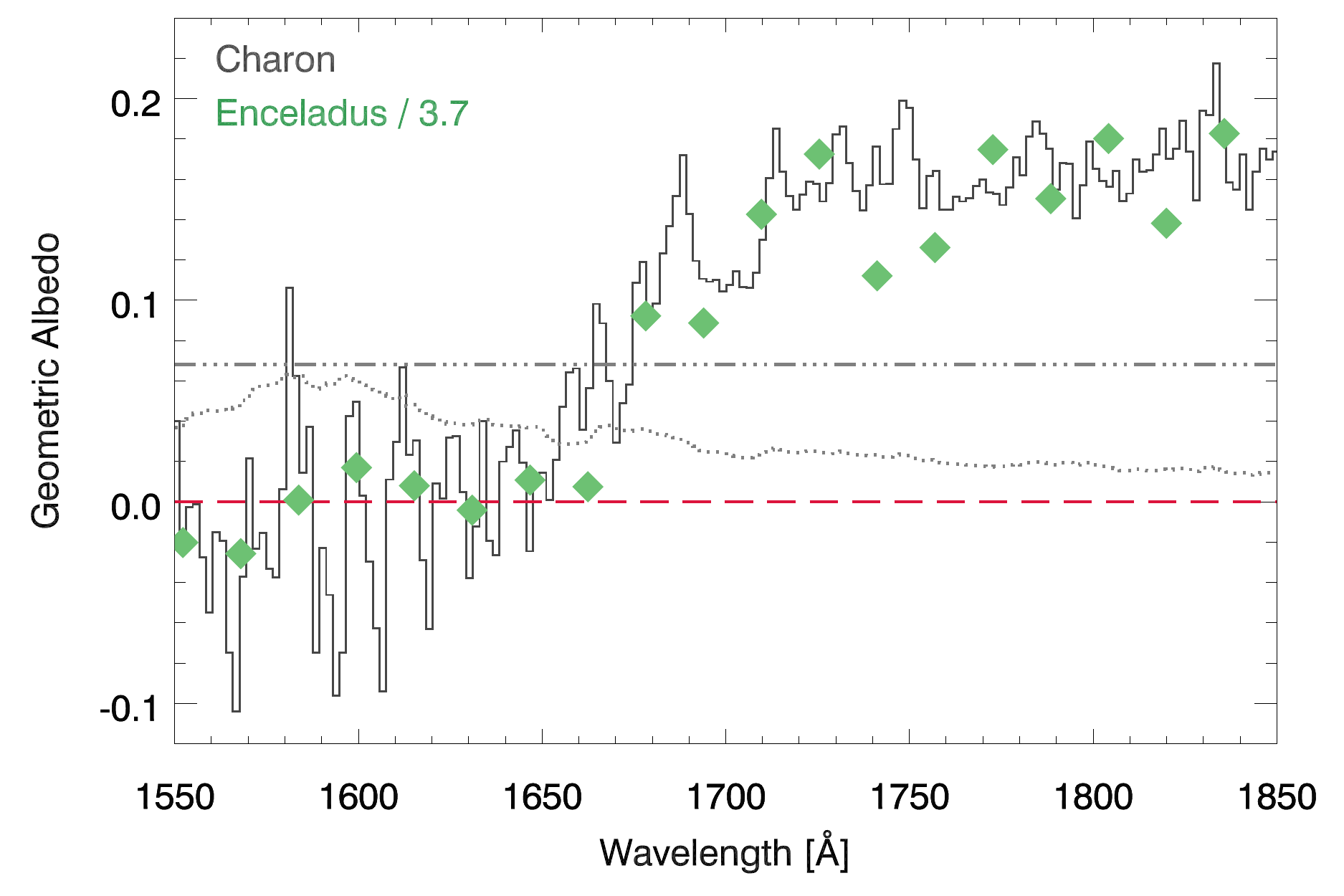}{\columnwidth}{}
  \vspace{-3em}
  \caption{FUV surface reflectance of Charon (black line) compared to Enceladus diamonds;][]{hendrix18}. As in \autoref{fig:albedo}, the dotted line shows the $1\sigma$ random uncertainty and the dot-dashed line shows the $1\sigma$ systematic uncertainty. When the geometric albedo of Enceladus is reduced by a factor of 3.7 it is very similar to Charon's. 
  \label{fig:enc}}
\end{figure}

\autoref{fig:enc} compares the geometric albedo of Charon as a function of wavelength to that of Enceladus using Cassini/UVIS data from \citet{hendrix18}. After reducing the geometric albedo of Enceladus by a factor of 3.7 (the hemispherically-averaged ratio of their albedos at 1800~\AA\ from \autoref{tab:alb}), we find that the FUV spectrum of Enceladus is a good match to Charon's, as the preceding comparison suggested.

\citet{hendrix10} found that the reflectance of Enceladus from the FUV through the visible could be explained if its surface is composed of primarily \ce{H2O} ice with a small amount of \ce{NH3} and a small amount of tholin. \NH\ observations of Charon in the visible and near-infrared find a very similar composition \citep{grundy16}; \ce{H2O} ice comprises most of the surface, but there are isolated areas of exposed \ce{NH3} ice or ammoniated hydrates, and Mordor Macula at its north pole is blanketed in red tholins. Thus, it is not surprising that the reflectances of Enceladus and Charon have similar FUV spectral shapes.

The tholins modeled by \citet{hendrix10} have wavelength-independent reflectances of $\sim0.05$ in the FUV and act as spectrally neutral darkening agents at these wavelengths. Further, other surface contaminants such as silicates, sulfur compounds, and organics that are not tholins can also be spectrally neutral in the FUV \citep[see, e.g.,][]{molyneux20}. Thus, the lower overall FUV reflectance of Charon compared to Enceladus could plausibly be explained by Charon having more and/or less-reflective surface contaminants. The presence of large, dark terrains on Charon (e.g., Mordor Macula) supports this hypothesis, as does Enceladus' much higher visible albedo (\autoref{tab:alb}).

\begin{figure}
  \fig{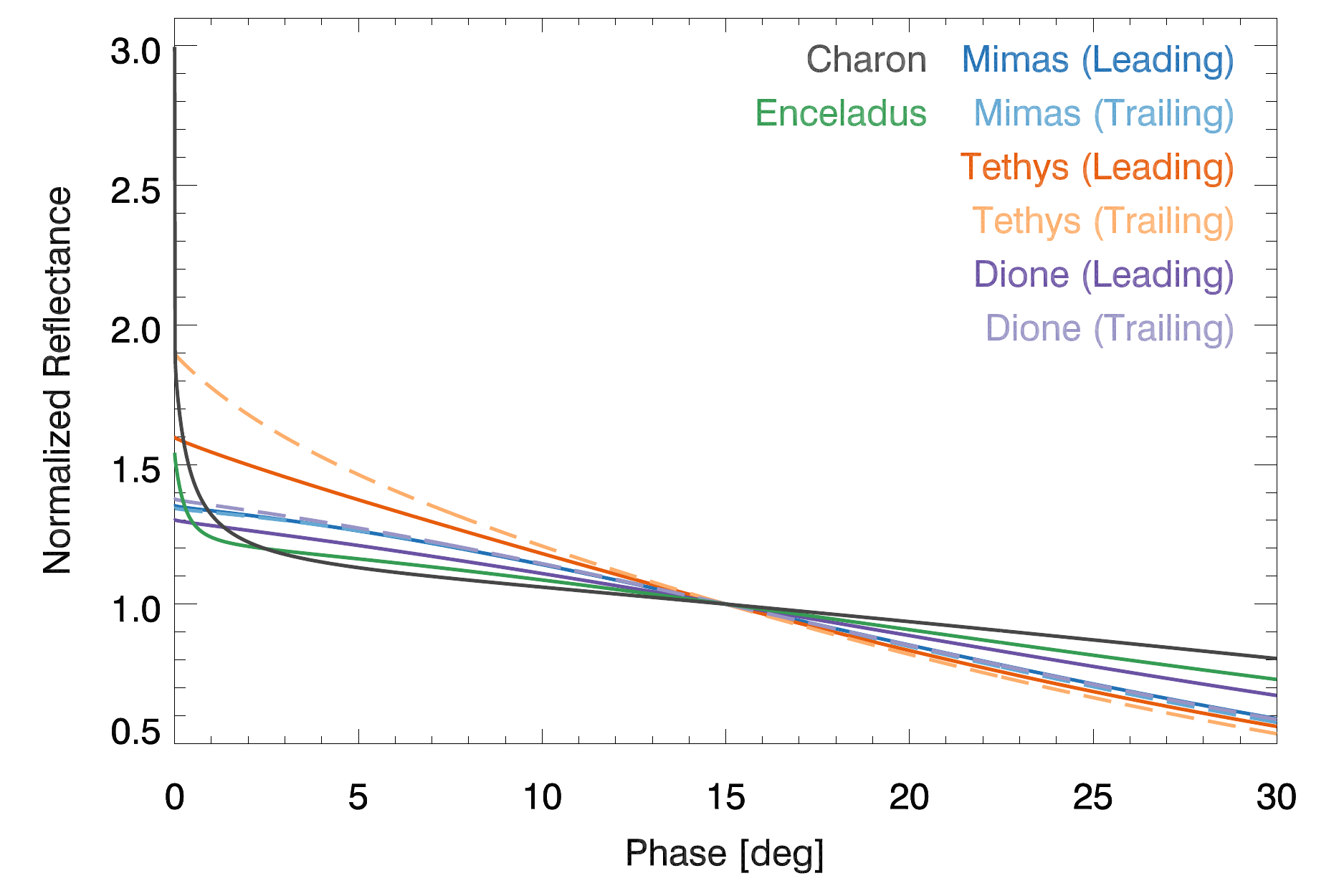}{\columnwidth}{}
  \vspace{-3em}
  \caption{Charon's solar phase curve (Verbiscer et~al. 2021, in preparation) compared to icy Saturnian satellites \citep{hendrix10,royer14} based on \citet{hapke12} model fits. The reflectance of all satellites has been set to unity at a phase angle of 15$\degr$ to emphasize variations in their solar phase behavior. 
    \label{fig:comp}}
\end{figure}

As discussed in \autoref{sec:disc:phase}, Charon's solar phase behavior differs from that of the Saturnian satellites. This is illustrated in \autoref{fig:comp}, which compares \citet{hapke12} model fits for Charon (Verbiscer et~al. 2021, in preparation), Enceladus \citep{hendrix10}, and the leading and trailing hemispheres of Mimas, Tethys, and Dione \citep{royer14} at solar phase angles of $0\degr$--$30\degr$. To emphasize the effects of solar phase variations at angles near where Charon was observed (see \autoref{fig:phase}), all these fits have been normalized to one at a solar phase angle of 15$\degr$.

Enceladus and the leading hemisphere of Dione (\autoref{sec:disc:phase}) exhibit the shallowest \replaced{decline}{declines} with increasing solar phase of the Saturnian satellites. Dione's trailing hemisphere and both hemispheres of Mimas have very similar solar phase behavior. Both hemispheres of Tethys decline somewhat more steeply than Mimas and Dione, as noted by \citet{royer14}.

Charon's brightness variation with solar phase is most simlar to Enceladus'. However, Charon has a stronger opposition surge (its brightness diminishes by $>50$\% from $0\degr$--$2\degr$) than Enceladus, followed by a shallower decline than any of the Saturnian satellites between phase angles of $15\degr$ and $30\degr$. We caution that Charon's opposition surge should only be directly compared to Enceladus' because Verbiscer et~al. (2021, in preparation) and \citet{hendrix10} incorporate both the shadow-hiding and coherent backscatter opposition effects in their \citet{hapke12} models, whereas \citet{royer14} only model the shadow-hiding opposition effect. It is beyond the scope of this work to determine why Charon's solar phase behavior differs from these Saturnian satellites, but we speculate that it is linked to their different physical environments.

Finally, we compare the FUV reflectances of Charon and Iapetus. Both satellites have similar sizes \citep[$\approx1500$~km diameter for Iapetus compared to $\approx1200$~km for Charon;][]{roatsch09,stern15}, and Iapetus' location well outside Saturn's E-ring yields a more benign environment than for the Saturnian satellites in \autoref{tab:alb}. The leading hemisphere of Iapetus is much darker than its trailing hemisphere, which has a very similar geometric albedo \citep[$\approx0.45$;][]{blackburn10} to Charon's in the visible. The geometric albedo of Iapetus in the FUV has not been published, but \citet{hendrix08} found that its bright terrain has $I/F\approx0.01$ at 1800~\AA\ and a solar phase angle of $\sim90\degr$. Using the scaled $V$-band phase curve of Verbiscer et~al. (2021, in preparation) to adjust Charon's 1800-\AA\ reflectance, we predict that Charon would have $I/F\sim0.016$ at the same solar phase. Thus, Charon's reflectance is $\sim50$\% larger than Iapetus' bright, trailing hemisphere in the FUV, even though Iapetus trailing hemisphere is slightly more reflective than Charon in the visible.

\section{Conclusions}
\label{sec:conc}

We have measured Charon's FUV surface reflectance for the first time by combining data from four observing sequences acquired by the Alice imaging spectrograph during the \NH\ flyby of the Pluto system. We find that Charon has the upturn in reflectance at 1650~\AA\ that is diagnostic of \ce{H2O} ice. This is reassuring since it has long been known that \ce{H2O} ice is the primary constituent of Charon's surface \citep{buie87}.

We find that the relative change in Charon's FUV reflectance with varying solar phase is well-described by its $V$-band phase curve (Verbiscer et~al. 2021, in preparation) scaled by a multiplicative factor of $\approx0.59$ (\autoref{fig:phase}). We do not observe rotational variability in Charon's FUV reflectance, at least over the small range of rotation we sampled. 

Charon has a geometric albedo of $<0.019$ ($3\sigma$) at 1600~\AA\ and $0.166\pm0.068$ at 1800~\AA\ (\autoref{fig:albedo}), which makes it darker in the FUV than the mid-UV \citep{krasnopolsky01,stern12}. We compare Charon's geometric albedo to that of the icy Saturnian satellites, and find that Charon and Enceladus have very similar spectral shapes in the FUV (\autoref{fig:enc}), although Enceladus is 3.7 times more reflective than Charon. We attribute the similarities in spectral shape to \ce{H2O} ice being the primary constituent of both bodies, and attribute the difference in absolute reflectance to a larger contribution of surface contaminants on Charon compared to Enceladus.

Finally, we compare Charon's solar phase behavior at 1800~\AA\ (\autoref{fig:comp}) to that of Enceladus, Mimas, Tethys, and Dione \citep{hendrix10,royer14}. Mimas and the trailing hemisphere of Dione have similar declines in reflectance between solar phases of $15\degr$ and $30\degr$, and Tethys declines more steeply. Enceladus' phase behavior is most similar to Charon's, but Charon exhibits a stronger opposition surge and a shallower decline from $15\degr$--$30\degr$.

\bigskip
We thank A.~Hendrix for Enceladus' FUV geometric albedo spectrum and helpful discussions. We also thank the Alice team, including our late team member David C. Slater, for an excellent instrument. This work was supported by NASA’s \NH\ project.

\bibliographystyle{aasjournal}
\bibliography{references}

\listofchanges

\end{document}